\documentclass{article}

\usepackage{amsmath,amsfonts}
\usepackage{algorithmic}
\usepackage{algorithm}
\usepackage{array}
\usepackage[caption=false,font=normalsize,labelfont=sf,textfont=sf]{subfig}
\usepackage{textcomp}
\usepackage{stfloats}
\usepackage{url}
\usepackage{verbatim}
\usepackage{graphicx}
\usepackage{cite}
\graphicspath{{images}}
\graphicspath{{./images/}}
\usepackage{xcolor}

\usepackage{hyperref}       
\usepackage{nicefrac}       
\usepackage{doi}
\usepackage{comment}
\usepackage[affil-it]{authblk}
\usepackage[font=footnotesize,labelfont=bf]{caption}

\providecommand{\keywords}[1]
{
  \small	
  \textbf{\textit{Keywords---}} #1
}

\begin{document}

\title{Modeling motor control in continuous-time\\ Active Inference: a survey}

\author[1,$\dagger$]{Matteo Priorelli}
\author[2,3,$\dagger$]{Federico Maggiore}
\author[2]{Antonella Maselli}
\author[2]{Francesco Donnarumma}
\author[2]{Domenico Maisto}
\author[2]{Francesco Mannella}
\author[1]{Ivilin Peev Stoianov}
\author[2]{Giovanni Pezzulo\thanks{Corresponding author: Giovanni Pezzulo, ISTC-CNR Via S. Martino della Battaglia 44, 00185 Rome, Italy, giovanni.pezzulo@istc.cnr.it}}

\affil[1]{\normalsize ISTC-CNR, National Research Council, Padua, Italy}
\affil[2]{\normalsize ISTC-CNR, National Research Council, Rome, Italy}
\affil[3]{\normalsize Department of Engineering, Roma Tre University, Rome, Italy}
\affil[$\dagger$]{\normalsize These two authors contributed equally}

\date{}

\maketitle

\begin{abstract}
The way the brain selects and controls actions is still widely debated. Mainstream approaches based on Optimal Control focus on stimulus-response mappings that optimize cost functions. Ideomotor theory and cybernetics propose a different perspective: they suggest that actions are selected and controlled by activating action effects and by continuously matching internal predictions with sensations. Active Inference offers a modern formulation of these ideas, in terms of inferential mechanisms and prediction-error-based control, which can be linked to neural mechanisms of living organisms. This article provides a technical illustration of Active Inference models in continuous time and a brief survey of Active Inference models that solve four kinds of control problems; namely, the control of goal-directed reaching movements, active sensing, the resolution of multisensory conflict during movement and the integration of decision-making and motor control. Crucially, in Active Inference, all these different facets of motor control emerge from the same optimization process - namely, the minimization of Free Energy - and do not require designing separate cost functions. Therefore, Active Inference provides a unitary perspective on various aspects of motor control that can inform both the study of biological control mechanisms and the design of artificial and robotic systems.
\end{abstract}



\section{Introduction}

A central question in computational motor control is how the brain selects and controls actions. A common assumption in formal frameworks such as Optimal Control \cite{tj02, dsi10} and reinforcement learning \cite{sb98} is that the building blocks of action control are stimulus-response mappings, or policies. Policies can be selected using either cheap-but-rigid, habitual mechanisms (i.e., based on the history of previous reinforcements), or costly-but-flexible, deliberative mechanisms (i.e., based on the value of action outcomes) \cite{dnd05}. Then, when a policy is selected, its execution follows stimulus-response control rules and could be accompanied by a process of prediction of action effects (via a so-called forward model) that helps manage delayed feedback \cite{wk98}.

An alternative to this stimulus-response perspective is the \emph{ideomotor} view that movements are selected and controlled on the basis of their effects (or outcomes), not of stimuli \cite{g70, h03, hmap01, l52, wgb03}. The general idea of the ideomotor theory is that the brain might learn the statistical (bidirectional) relations between actions and their effects and then use the learned action-effect codes both to predict action consequences (in the forward direction, from actions to effects) and to select and plan actions that achieve the intended effects (in the backward direction, from effects to actions). Stimuli could be part of this picture, leading to stimulus-action-effect codes, but they would not be the main responsible for action planning, selection, and control. Various empirical findings support the claim of ideomotor theory that action effects influence the selection and control of actions. For example, one study required participants to press one of four horizontally arranged buttons in response to \emph{color} stimuli \cite{k01}. After each keypress, an \emph{effect} stimulus was shown in one of four horizontally arranged locations. Crucially, responses to \emph{color} stimuli were faster when there was a correspondence between the locations of the \emph{effect} stimulus and the pressed button. This occurred despite color stimuli appeared before effect stimuli, indicating that anticipated action effects influenced actions. This influence would not be present if action selection was stimulus-response, because effects only occur after actions are completed. Other subsequent studies suggest that the influence of effects over actions regards various processes, such as action planning, selection, preparation, initiation, and control \cite{a02, h96, kkh04, mwehg08} and they could already be present in infancy \cite{pheb12}.

Another framework that highlights the centrality of action effects (or outcomes) rather than stimulus-response codes is cybernetics \cite{w48}. For example, the early TOTE (test, operate, test, exit) model assumes that the brain continuously tests whether there is a mismatch between an internally defined event (roughly, a goal, outcome, or setpoint) and the currently sensed event; and in case it detects a mismatch, it triggers a corrective action that reduces it \cite{mgp60}. A simple illustration of this ``closed-loop" control mechanism is the functioning of a thermostat: a discrepancy between a desired temperature or set point (say, 37 degrees) and a sensed temperature (say, 35 degrees) triggers an action (say, switch a heating device on) until there is no more mismatch. While the TOTE scheme did not natively include sophisticated planning or control mechanisms, it exemplifies a mechanism by which it is an internal matching operation, not a stimulus, which triggers actions. Similarly, in perceptual control theory, the central goal of a system is to continuously monitor that some internally represented perceptual variable has the desired value (e.g., that the number indicated by a speedometer in a car is 80 km/h) and if not, trigger corrective actions (accelerate or decelerate) that cancel out discrepancies from the desired value \cite{m11, pp}.



The two above (ideomotor and cybernetic) schemes require the brain to internally represent action effects and other expected (or intended) events and to continuously perform matching operations to calculate discrepancies between expected and sensed events. Recently, these internal matching operations and discrepancies have become the focus of a large set of theoretical and empirical studies \cite{cp16, h13, ppf22, p17, wmco20}, which are typically conceptualized in terms of predictive coding and Active Inference theories \cite{ppf22}. These theories assume that the brain maintains a statistical model of the regularities of the environment and uses it to continuously generate predictions about present and future events, including action effects. 
Crucially, the brain models can include some preferred states (e.g., desired values of physiological parameters or setpoints for motor control) that regulate the control of externally directed actions such as overt movements \cite{asf12} and internal regulatory actions \cite{prf15, tbmbsp21}. 
This is because the brain continuously predicts the desired values (e.g., of body posture or temperature) and monitors discrepancies with sensed stimuli. 
Any discrepancy is registered as a prediction error that triggers a corrective action that minimizes the error (or alternatively, depending on the context, leads to model revision and learning) that ultimately ensures that the system remains within its preferred states. 

In keeping with ideomotor and cybernetics accounts of motor behavior, Active Inference formalizes the problem of motor control by assuming that agents act on the surrounding environment in a goal-directed manner, to achieve a desired state. Active Inference agents monitor the state of the system (which may include the external environment and their own bodily configuration) through the senses (i.e., perception) and continuously predict how the state of the system will evolve in time. 
This predictive processing is granted by an internal representation of the system dynamics, which is assumed to be learned through exposure to the statistical regularities that govern the environment and the body (the laws of physics, kinematic regularities, etc.), both during the lifespan and throughout evolution. 
Furthermore, Active Inference agents continuously formulate sensory predictions (e.g., about action outcomes) and compare them with sensory events gathered through the senses. 
The resulting sensory prediction errors are considered within the state estimation process, which strives to minimize the errors. 
To minimize the prediction errors, the (brain of the) agent has two ways. 
First, it can change the model that generated the predictions in the first place: this amounts to a process of belief revision and learning in theories of predictive coding, DEM and generalized filtering \cite{f05, rb99}, \cite{e23101306, fsld10}. Second, the agent can minimize prediction errors by acting upon the system and changing its state, in such a way that the system-produced events (registered as sensory events by the agent) become more similar to its sensory predictions. This second way to minimize prediction errors -- by acting -- is key to Active Inference and permits formalizing early ideomotor and cybernetic ideas, by linking them to the biologically plausible scheme of predictive coding \cite{ppf22}.

While several surveys and tutorials can be found in the literature about Active Inference in discrete state spaces \cite{DaCosta2020, Smith2022}, the continuous framework has received relatively less attention. Discrete models are critical to perform planning and decision-making, but handling continuous signals is key when interacting with the external environment. Given the growing interest in predictive processing, here we provide an overview of how motor control is modeled.


In the rest of this article, we provide a short formal introduction to Active Inference; we discuss various examples of Active Inference models of motor control; finally, we discuss the unique features of this framework.


\section{Active Inference in continuous time}

Active Inference has been used to model a large variety of problems of motor control, decision-making, planning and rule learning that are relevant for both biological organisms and robots \cite{asf12, dcafp17, mlp22, pnfp16, a23}. This section provides a concise formal introduction to Active Inference in continuous time; a more detailed treatment comprising discrete-time formulations can be found in \cite{ppf22}.

Active Inference is built upon the Free Energy Principle (FEP), which assumes that all living organisms strive to minimize the “surprise associated with sensory exchanges with the world”, allowing them to resist a natural tendency to disorder \cite{ppf22}. The Variational Free Energy (VFE) -- or Free Energy, for brevity -- $\mathcal{F}$ is introduced as a mathematically treatable upper bound on surprise; it is a functional that is widely used in statistics as part of Variational Bayes methods \cite{b15}, and it is analogous to the evidence lower bound (ELBO) used in machine learning. Active Inference appeals to the minimization of Free Energy to model the action-perception loop of living organisms and assumes that both action and perception minimize the same (Free Energy) quantity, as will become clear later.

\begin{figure*}[!t]
    \centering
    \includegraphics[width=0.8\textwidth]{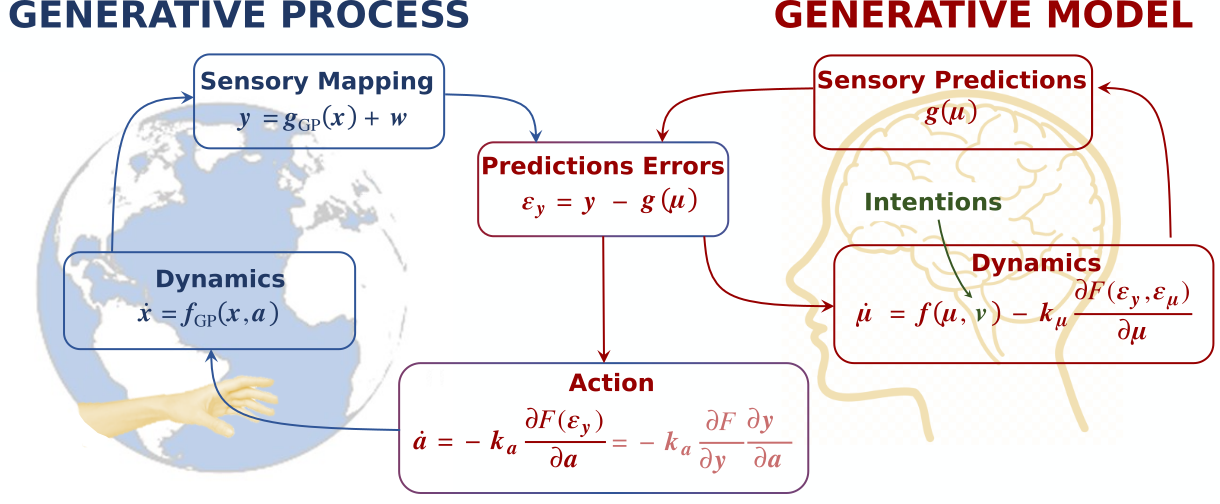}
    \caption{Generative process (in blue) and generative model (in red) in Active inference. The generative process includes the dynamics $\boldsymbol{f}_{GP}(\boldsymbol{x},\boldsymbol{a})$ that governs the temporal evolution of the hidden states $\boldsymbol{x}$, and the sensory mapping $\boldsymbol{g}_{GP}(\boldsymbol{x})$ that maps the hidden states into the sensory input $\boldsymbol{y}$ that the agent receives from the environment, along with sensory noise $\boldsymbol{w}$. The generative model includes an internal representation $\boldsymbol{f}$ about the system dynamics generated from a belief $\boldsymbol{\mu}$ over the hidden states, and an internal model $\boldsymbol{g}$ that maps this belief into sensory predictions. The latter are used to compute prediction errors $\boldsymbol{\varepsilon}_y$ which contribute to perceptual inference and the generation of actions $\boldsymbol{a}$ via the corresponding Free Energy minimization terms. Eventually, actions affect the generative process by entering the system dynamics. Note that $\boldsymbol{f}$ does not include an explicit representation of the actions and is instead driven by internal representations of intentions encoded into the hidden causes $\boldsymbol{\nu}$.}
    \label{fig1}
\end{figure*}

Any implementation of an Active Inference agent requires specifying two interacting systems, as shown in Fig. \ref{fig1}. The first one is the ``generative process”, which describes how the physical system which the agent interacts with (e.g., the environment and/or the agent's body) evolves in time, and how it maps into the sensory inputs observed by the agent. The second one is the “generative model”, which describes the internal model that the agent holds about how the system is expected to evolve in time and to map into sensory states. As illustrated in Fig. \ref{fig1}, the two systems interact bidirectionally: the generative process determines the sensory inputs that the agent receives and processes (e.g., to compute prediction errors), while the generative model produces actions that influence the dynamics of the generative process.

The action-perception loop of Active Inference can be summarized as follows. Consider an agent immersed in a dynamic environment and receiving observations $\boldsymbol{y}$ generated from hidden variables $\boldsymbol{u}$, which generally consist of hidden states $\boldsymbol{x}$ and hidden causes $\boldsymbol{v}$ (but they could also include other variables, like parameters evolving on different time scales) \cite{f08}.
The variational method approximates the intractable posterior $P(\boldsymbol{u} | \boldsymbol{y}) = P(\boldsymbol{u}, \boldsymbol{y}) / P(\boldsymbol{y})$ through the definition of an auxiliary, approximate posterior distribution $Q(\boldsymbol{u})$, sometimes called recognition density \cite{b15,bkms17}. 
The approximation is achieved by minimizing the Kullback-Leibler (KL) divergence between these two distributions. However, since this quantity still depends on the intractable marginal $P(\bf{y})$, it is replaced by the (formally equivalent) minimization of VFE $\mathcal{F}$. The latter provides an upper bound on log evidence (or surprise) and its minimization is tractable because it depends on two quantities that the agent knows or has inferred: the joint probability $P(\boldsymbol{u}, \boldsymbol{y})$ and the approximate posterior $Q(\boldsymbol{u})$.

Since the optimization of $\mathcal{F}$ for an arbitrary $Q(\boldsymbol{u})$ is often complex, it is common to make some additional (biologically plausible) assumptions. 
A standard assumption made in Active Inference is the Laplace approximation \cite{fmtap07}, which implies that the approximate posterior is a multivariate Gaussian distribution, i.e., in the simple case of $\boldsymbol{u}=\{ \boldsymbol{x}, \boldsymbol{v} \}$, $Q(\boldsymbol{u}) = \mathcal{N}(\{ \boldsymbol{\mu}, \boldsymbol{\nu} \},\boldsymbol{\Pi}^{-1})$, where $\boldsymbol{\mu}$ is the best guess or belief about the hidden states, $\boldsymbol{\nu}$ is the belief about the hidden causes, and $\boldsymbol{\Pi}$ is their precision or inverse covariance matrix. 
A second common assumption used to simplify the recognition density is the mean-field approximation \cite{f08}, which renders some variables of the model conditionally independent (for example, hidden state variables and other model parameters that we did not consider here).
Given the above assumptions, it is possible to transform the functional $\mathcal{F}$ into a function and evaluate it, up to constant quantities that do not affect the minimization process: 
\begin{equation}
    \mathcal{F} \approx - \ln P(\boldsymbol{u}, \boldsymbol{y}) |_{\boldsymbol{x} = \boldsymbol{\mu}, \boldsymbol{v} = \boldsymbol{\nu}}
\end{equation}

Hence, under the FEP, everything is reduced to a process of Free Energy minimization; however, this requires specifying a generative model - which is our next topic.


\subsection{Generative model}

Designing the generative model implies additional assumptions regarding how an agent represents the system dynamics and the mapping into its sensory inputs. 
A generative model can be described in terms of a joint probability density $P(\boldsymbol{u}, \boldsymbol{y}) = P(\boldsymbol{y} | \boldsymbol{u}) P(\boldsymbol{u})$, which highlights the separation between two components: the observation (or likelihood) model and the prior about the hidden variables. The latter can be further factorized into the joint density $P(\boldsymbol{u}) = P(\boldsymbol{x}|\boldsymbol{v})P(\boldsymbol{v})$. 
In general, the hidden causes $\boldsymbol{v}$ are quantities that act as causal variables (or priors) over the hidden states $\boldsymbol{x}$ used to describe the environment, thus enriching the representation of the system dynamics.
In some Active Inference implementations, hidden causes are used to encode the agent’s goals (as will become clear when we discuss specific examples). This is because, in keeping with ideomotor and cybernetic formulations, any deviation from the hidden causes is registered as a prediction error that the agent tries to minimize.

The dynamic environment represented by an agent is usually modeled with the following stochastic equations:
\begin{equation}
    \boldsymbol{y} = \boldsymbol{g}(\boldsymbol{x}, \boldsymbol{v})  + \boldsymbol{\omega}_y \quad \dot{\boldsymbol{x}} = \boldsymbol{f}(\boldsymbol{x}, \boldsymbol{v})  + \boldsymbol{\omega}_x \quad \boldsymbol{v} = \boldsymbol{\eta}  + \boldsymbol{\omega}_v
\end{equation}
where the function $\boldsymbol{g}$ converts latent variables $\boldsymbol{x}$ and $\boldsymbol{v}$ into observed states $\boldsymbol{y}$, $\boldsymbol{f}$ encodes the evolution of the hidden states over time, $\boldsymbol{\eta}$ is the mean of the prior distribution over the hidden causes, while $\boldsymbol{w}_y$, $\boldsymbol{w}_x$ and $\boldsymbol{w}_v$ are noise terms describing system uncertainty, here assumed to belong to multivariate normal distributions with zero mean and precisions $\boldsymbol{\Pi}_y$, $\boldsymbol{\Pi}_x$, and $\boldsymbol{\Pi}_v$.

This leads to the VFE:
\begin{align}
\begin{split}
    \mathcal F &\approx \frac{1}{2} \left[ \boldsymbol{\varepsilon}_y^T \boldsymbol{\Pi}_y \boldsymbol{\varepsilon}_y + \boldsymbol{\varepsilon}_x^T \boldsymbol{\Pi}_x \boldsymbol{\varepsilon}_x + \boldsymbol{\varepsilon}_v^T \boldsymbol{\Pi}_v \boldsymbol{\varepsilon}_v \right] \\
    \boldsymbol{\varepsilon}_y &= \boldsymbol{y} - \boldsymbol{g}(\boldsymbol{\mu}, \boldsymbol{\nu}) \\
    \boldsymbol{\varepsilon}_x &= \boldsymbol{\mu'} - \boldsymbol{f}(\boldsymbol{\mu}, \boldsymbol{\nu}) \\
    \boldsymbol{\varepsilon}_v &= \boldsymbol{\nu} - \boldsymbol{\eta}
\end{split}
\end{align}
where $\boldsymbol{\mu}$ and $\boldsymbol{\mu'}$ are respectively the internal representations (beliefs) about the 0th and 1st temporal orders of the hidden states $\boldsymbol{x}$ and $\dot{\boldsymbol{x}}$. Hence, the VFE takes the form of a sum of quadratic forms of prediction errors: a sensory prediction error $\boldsymbol{\varepsilon}_y$, a state or model prediction error $\boldsymbol{\varepsilon}_x$, and a prior prediction error $\boldsymbol{\varepsilon}_v$.

To effectively represent complex dynamics of the generative process, it is possible to improve the agent's model by using generalized coordinates of motion \cite{fsld10, ftd08} beyond the 1st order. For instance, suppose that the brain represents beliefs about the position of an object. Under a generalized coordinates model, it would also maintain beliefs about its velocity, acceleration, jerk, and so on. All these time derivatives are then concatenated to form a generalized belief, denoted by a vector $\tilde{\boldsymbol{\mu}} \equiv \left[ \boldsymbol{\mu}, \boldsymbol{\mu}', \boldsymbol{\mu}'', \, \dots \right]$. The same notation is used for the time derivatives of the other variables (i.e., $\boldsymbol{\tilde{y}} \equiv \left[ \boldsymbol{y}, \boldsymbol{y}', \boldsymbol{y}'', \, \dots \right] $).

Applying a local linearization \cite{ftd08} to the system dynamics, and then eliminating the cross terms in the derivatives, it is possible to express the generative model by the following set of equations:
\begin{equation}
\begin{split}
    \boldsymbol{y} &= \boldsymbol{g}(\boldsymbol{x})  + \boldsymbol{\omega}_y \\
    \boldsymbol{y}' &= \frac{\partial \boldsymbol{g}(\boldsymbol{x})}{\partial \boldsymbol{x}} \boldsymbol{x}'  + \boldsymbol{\omega}'_y  \\
    \boldsymbol{y}'' &= \frac{\partial \boldsymbol{g}(\boldsymbol{x})}{\partial \boldsymbol{x}} \boldsymbol{x}'' +\boldsymbol{\omega}''_y \\
    & \qquad \vdots
\end{split}
\qquad
\begin{split}
    \boldsymbol{x}' &= \boldsymbol{f}(\boldsymbol{x}, \boldsymbol{v})  + \boldsymbol{\omega}_x \\
    \boldsymbol{x}'' &= \frac{\partial \boldsymbol{f}(\boldsymbol{x}, \boldsymbol{v})}{\partial \boldsymbol{x}}\boldsymbol{x}' + \boldsymbol{\omega}'_x    \\
    \boldsymbol{x}''' &= \frac{\partial \boldsymbol{f}(\boldsymbol{x},  \boldsymbol{v})}{\partial \boldsymbol{x}} \boldsymbol{x}'' + \boldsymbol{\omega}''_x \\
    & \qquad \vdots
\end{split}
\end{equation}
which can be expressed in a compact form as $\tilde{\boldsymbol{y}} = \tilde{\boldsymbol{g}}(\tilde{\boldsymbol{x}}) + \tilde{\boldsymbol{\omega}}_y$ and $\mathcal{D} \tilde{\boldsymbol{x}} = \tilde{\boldsymbol{f}}(\tilde{\boldsymbol{x}}, \boldsymbol{v}) + \tilde{\boldsymbol{\omega}}_x$.
Here, the $\mathcal{D}$ operator maps each element of the generalized coordinates to its time derivative: $ \mathcal{D} \tilde{\boldsymbol{\mu}} = \left[ \boldsymbol{\mu}', \boldsymbol{\mu}'', \boldsymbol{\mu}''', \, \dots \right] $.
Note that using generalized coordinates permits dealing not only with white but also colored noise. In general, these coordinates have been introduced to deal with non-Markovian processes \cite{GVK021832242, f08}, adopting a Stratonivich interpretation with continuous stochastic variables having finite, non-zero autocorrelation functions. This is usually done using a temporal covariance matrix acting as a Gaussian filter between noise terms that modify generalized precision matrices, denoted as $\boldsymbol{\tilde{\Pi}}$. This leads to a Free Energy that has the same (quadratic) form of prediction errors \cite{ftd08}.


\subsection{Free Energy minimization}

Active Inference assumes that the Free Energy is minimized in two complementary ways. One, shared with predictive coding \cite{rb99}, consists of modifying the agent’s internal beliefs, to produce predictions that match the current observations. In particular, it has been proposed \cite{f08} that the intrinsic dynamics of neural activity evolve in such a way as to implement a (modified) gradient descent scheme:
\begin{align}
\label{eq:descent}
    \dot{\tilde{\boldsymbol{\mu}}} - \mathcal{D}\tilde{\boldsymbol{\mu}} &= -k_{\mu} \frac{\partial \mathcal{F}}{\partial \tilde{\boldsymbol{\mu}}} &
    \dot{\boldsymbol{\nu}} &= - k_{\nu}\frac{\partial \mathcal{F}}{\partial \boldsymbol{\nu} }
\end{align}
where $k_{\mu}$ and $k_{\nu}$ are tunable learning rates.

In biological terms, this means that agents are constantly engaged in an inferential process to capture the hierarchical relationships between what is perceived at every instant and what causes those perceptions. Importantly, the instant derivative of a particular order of the generalized belief does not necessarily correspond to the belief over that derivative (i.e., $\dot{\boldsymbol{\mu}} \neq \boldsymbol{\mu}'$); the difference between those two terms provides an additional error-term to minimize. As evident in Eq. \ref{eq:descent}, it is only when the Free Energy is minimized that the generalized belief captures the real instantaneous trajectory of the environment. Furthermore, since the Free Energy consists of a sum of quadratic forms, its partial derivatives lead to simple update equations -- proportional to the prediction errors weighted by their respective precisions -- which are similar to those originally derived in the predictive coding model of \cite{rb99}:

\begin{equation}
    \dot{\tilde{\boldsymbol{\mu}}} \propto \partial_{\tilde{\mu}} \tilde{\boldsymbol{g}}^T \tilde{\boldsymbol{\Pi}}_y \tilde{\boldsymbol{\varepsilon}}_y + \partial_{\tilde{\mu}} \tilde{\boldsymbol{f}}^T \tilde{\boldsymbol{\Pi}}_x \tilde{\boldsymbol{\varepsilon}}_x - \mathcal{D}^T \tilde{\boldsymbol{\Pi}}_x \tilde{\boldsymbol{\varepsilon}}_x
\end{equation}

Thus, the overall update for the generalized belief over hidden states is subject to three different forces: a likelihood component proportional to the sensory prediction error; forward and backward components of the state prediction errors coming from the previous and next temporal orders.

In Active Inference, there is however a second way to minimize the Free Energy: by acting in the environment, the agent produces sensory observations that match its predictions. This action-related way to minimize the Free Energy is especially appealing to model biological organisms that strive to realize their goals (or the prior preferences encoded in their generative models). For example, if an organism is endowed with a prior over a desired body temperature and senses a different value, it can maintain its integrity by acting (e.g., by moving to a place having the desired temperature) -- whereas only changing its belief would probably lead to death in the long term. In short, the predictions are fulfilled by acting, rather than corrected by changing mind.

\begin{figure*}[h]
    \centering
    \includegraphics[width=\textwidth]{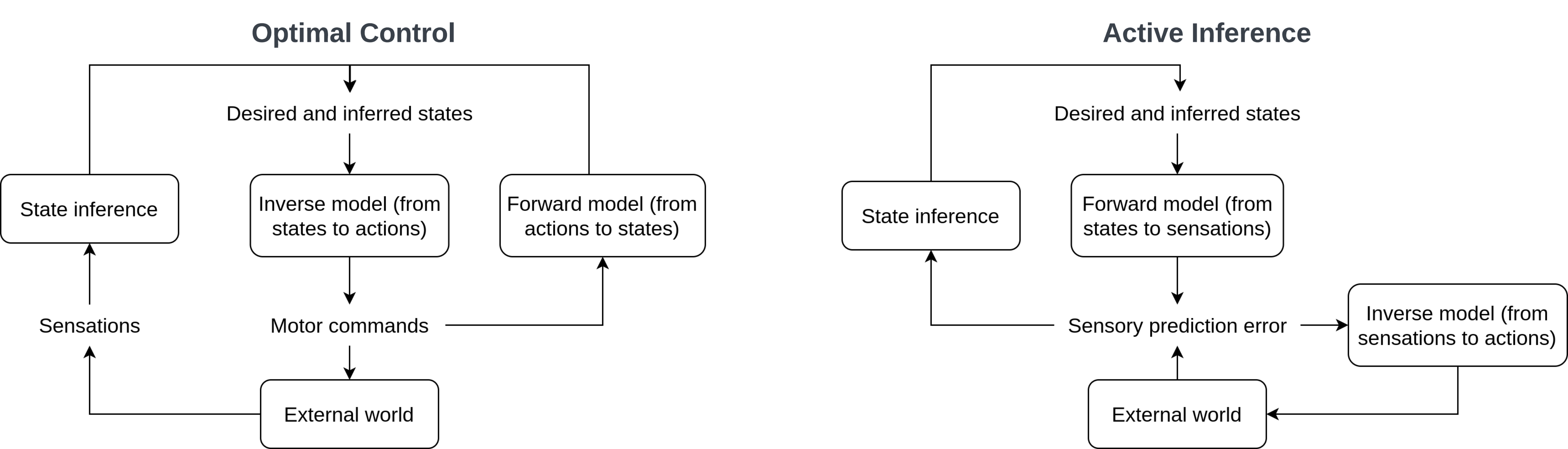}
    \caption{Comparison between motor control schemes in Optimal Control (left) and Active Inference (right). See the main text for a detailed discussion.}
    \label{fig2}
\end{figure*}

Formally, minimizing the Free Energy with respect to the actions $\boldsymbol{a}$ results in a gradient descent scheme similar to the previous case:
\begin{equation}
\label{eq:action}
    \dot{\boldsymbol{a}} = - k_{a}  \frac{\partial \mathcal{F} }{\partial \boldsymbol{a}} = - k_{a}  \frac{\partial \mathcal{F} }{\partial \boldsymbol{y}} \frac{\partial \boldsymbol{y}}{\partial \boldsymbol{a}}
\end{equation}
where $k_{a}$ is a learning rate.

Note that in the last gradient of Eq. \ref{eq:action}, the presence of the term $\partial_a \boldsymbol{y}$ is key -- and points to the agent's ``implicit'' knowledge of (simple) sensory outcomes of actions. Here, ``implicit" is used in the sense that the action variable $\boldsymbol{a}$ is not considered to be part of the generative model, but at the interface between generative model and process (see Fig. \ref{fig1}). This knowledge does not correspond to a sophisticated ``inverse" model, but to simple and short-term action consequences, which could be associated with reflex arcs. For example, the ``inverse" model for a velocity-controlled scheme could be simply approximated by a time constant $\Delta_t$ \cite{olc22}. As will be explained below, in the Active Inference framework reflex arcs are key to motor execution via the minimization of proprioceptive prediction errors induced by top-down modulatory signals from motor areas \cite{asf12}. 

In sum, the above discussion highlights that the interaction between the agent and the environment is characterized by a closed loop, during which the agent minimizes the Free Energy (or under some simplifying assumptions, prediction errors). The agent could minimize prediction errors through belief updates, which renders the generative model closer to the generative process, therefore creating a good representation of the environment. Alternatively, it could generate actions that render the generative process closer to the generative model. Whether the former (belief updating) or the latter (action) process is selected simply depends on the relative balance between prediction errors and their relative precisions. For example, an agent endowed with an extremely precise prior would never update it in the light of novel evidence and hence would always try to minimize the Free Energy by acting. Conversely, an agent endowed with an imprecise prior would be more willing to update its beliefs in the light of novel evidence. This implies that the design of the generative model is a crucial choice to determine the agent's behavior. Two agents that deal with the same situation but are endowed with different generative models (e.g., with different priors) could produce completely different patterns of behavior. 


\subsection{Neural underpinnings of motor control in Active Inference}

It is useful to briefly summarize the key biological assumptions of motor control in Active Inference and compare them with classical theories such as Optimal Control; see \cite{f11} for an extensive treatment of these differences and of forward and inverse models in the brain. As we highlight in Fig. \ref{fig2}, both Active Inference and Optimal Control \cite{tj02, dsi10} use similar processes and variables, but arrange them differently. The main difference is that in Active Inference, the forward (generative) models convey proprioceptive predictions down to the spinal cord to compute motor commands at the level of the reflex arcs. Instead, in Optimal Control, the forward models are coupled with inverse models at a higher level to compute motor control signals.

Thus, in Active Inference action is made possible through low-level suppression of prediction errors, which not only climb up the hierarchy but also exert forces on the muscle states. 
It is worth noting that while the mathematical formulation of Active Inference allows involving any sensory modality in action processes, in biological treatments it is often assumed that motor control is realized by minimizing proprioceptive - and not exteroceptive - prediction errors \cite{asf12}, whereas autonomic control is realized by minimizing interoceptive prediction errors \cite{prf15, sf16}. 
The reason for assuming that only proprioceptive prediction errors are involved in motor control is the observation that the efferents of the somatomotor system share crucial similarities with top-down projections of other brain areas, thus seeming to encode proprioceptive predictions rather than motor commands \cite{asf12}. 
In this perspective, proprioceptive prediction errors are computed in the spinal cord through the reflex arcs and backpropagated by afferents throughout the cortical hierarchy, whereas exteroceptive prediction errors are generated locally in their respective functional areas. 
It is thus unlikely that direct somatomotor efferents convey pure exteroceptive signals from the respective sensory areas to the muscles, or that a difficult inversion is realized in the spinal cord between motor and exteroceptive domains. This makes it unlikely that exteroceptive sensations are directly used for action execution and marks another significant difference with Optimal Control. However, movements driven only by proprioceptive contributions raise a few concerns regarding multisensory conflict resolution, as will be explained later.


\section{Examples of Active Inference models of motor control}

Here we review some examples of Active Inference models that target four kinds of problems: goal-directed motor control, active sensing, multisensory conflict resolution, and decision-making in dynamic environments. The list of selected models and functions is not exhaustive, but provides an overview of the scope of Active Inference in continuous time.


\subsection{Goal-directed reaching: four examples}

\begin{figure*}[!t]
    \centering
    \includegraphics[width=1.0\textwidth]{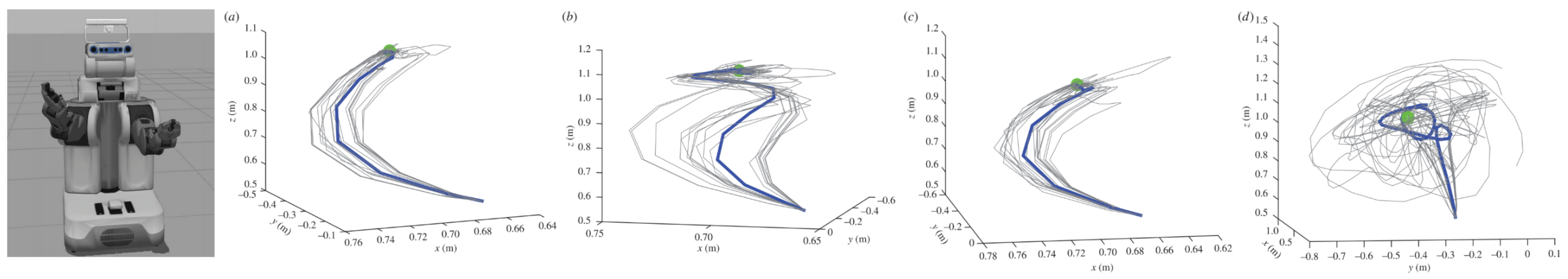}
    \caption{An example of goal-directed reaching, from \cite{pnfp16}. Left panel: the stimulated robot. Right panels: Reaching trajectories of the robot's 7 DoF arm in three dimensions, from a start location (bottom) to a goal location (green dot) in four scenarios: (subpanel a) without noise; (subpanel b) with noisy proprioception; (subpanel c) with noisy vision; (subpanel d) with noisy proprioception and vision. The blue trajectory is the mean of the 20 trajectories, shown in gray.
}
    \label{fig3}
\end{figure*}

As explained above, action follows prediction errors that may result from a discrepancy between a proprioceptive observation and a proprioceptive prediction. This situation can be exemplified in the case of goal-directed reaching actions, where the start and the goal positions of the agent's arm initially differ. One example of Active Inference implementation of a goal-directed reaching task was proposed in \cite{pnfp16}. Here, a simulated 7 Degrees of Freedom (DoF) robotic arm had to reach a static target (see Fig. \ref{fig3}, left panel). The agent maintained a belief over the arm’s joint angles, and was endowed with a proprioceptive model producing predictions in the same domain and a visual model generating the end effector’s position. The goal states were embedded into the 1st-order dynamics function, where the trajectory of each joint was modeled by Newtonian dynamics with parameters $\boldsymbol{\lambda}$, $\boldsymbol{\kappa}$ and $\boldsymbol{m}$ representing elasticity, viscosity, and mass. The role of the dynamics function is to make the agent perceive a force proportional to the desired one, (or better, ``think that it will perceive" the force, since the force does not have any counterpart in the generative process). This force is computed by performing a kinematic inversion of the error between the end effector and the target's position, and by taking into account all the possible singularities. The right panels of Fig. \ref{fig3} show the different trajectories that the agent performs under various conditions in which the noise of one or more information sources is introduced. These results highlight the advantages of relying on precise multisensory information (panel a) compared to situations in which proprioceptive (panel b), visual (panel c), or both sources (panel d) are noisy.

\begin{figure}[!t]
    \centering
    \includegraphics[width=0.8\textwidth]{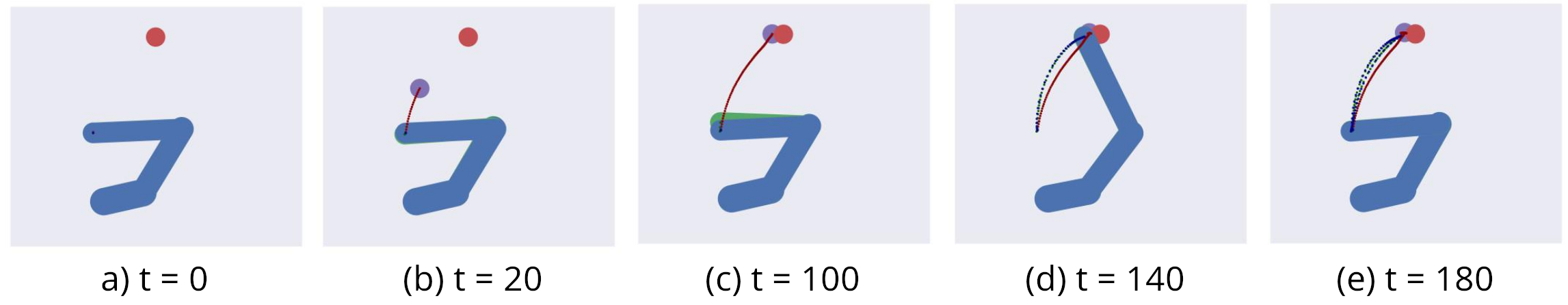}
    \caption{Visual representation of the two-step delayed reaching task of \cite{ps22a}, composed of two distinct phases. Real and estimated arms are displayed in blue and green, real and estimated targets in red and purple.}
    \label{fig_arm}        
\end{figure}

Another implementation for a similar reaching task is described in \cite{ps22a}. Here, the agent had to perform two tasks: continuously tracking a moving target and realizing multi-step movements toward two different object locations. In this case, the high-level belief consisted not only of the arm’s joint angles, but also of as many components as every object to interact with. Such components were encoded in the proprioceptive domain and they could be interpreted as particular affordances that the agent wanted to realize. To address the first (target tracking) task, the belief dynamics uses a custom ``intention" that manipulates the current belief to produce a possible future configuration. For example, if the belief consists of three components corresponding to joint configurations of an arm, a target, and a previously memorized home button - i.e., $\boldsymbol{\mu} = [\boldsymbol{\mu}_a, \boldsymbol{\mu}_t, \boldsymbol{\mu}_h]$ - an intention to reach the target is built through a function that sets the first component equal to the second one, i.e., $\boldsymbol{\mu}_t^* = [\boldsymbol{\mu}_t, \boldsymbol{\mu}_t, \boldsymbol{\mu}_h]$. This future prediction is then subtracted from the current belief and embedded into the 0th-order dynamics:
\begin{equation}
    f^{(t)} = \lambda (\boldsymbol{\mu}_t^* - \boldsymbol{\mu}) = \lambda \cdot [\boldsymbol{\mu}_t - \boldsymbol{\mu}_a, \boldsymbol{0}, \boldsymbol{0}]
\end{equation}
where $\lambda$ is an attractive gain. Since the target configuration is continuously inferred through visual predictions (here, generated by the decoder of a Variational Autoencoder (VAE) \cite{kw13}, the agent is able to reach and track moving objects, as shown in Fig. \ref{fig4}. Furthermore, \cite{ps22a} generalized the above approach by considering multiple intentions that operate simultaneously, thus allowing one to realize more complex movements or multi-step tasks, such as the one represented in Fig. \ref{fig_arm}. For example, if a home button reaching intention is constructed in the same way as the first one, called $\boldsymbol{\mu}_h^*$, the overall attractive force is:
\begin{equation}
    \dot{\boldsymbol{\mu}}' = -\pi_{x,t} \boldsymbol{\varepsilon}_{x,t} - \pi_{x,h} \boldsymbol{\varepsilon}_{x,h}
\end{equation}
where $\boldsymbol{\varepsilon}_{x,t}$ and $\boldsymbol{\varepsilon}_{x,h}$ are the state prediction errors of the two intentions, with precisions $\pi_{x,t}$ and $\pi_{x,h}$. A multi-step behavior may then be achieved by dynamically modulating the latter, e.g., through a belief over tactile sensations \cite{ps22b}. Note that while the above examples suggest that the presence of a goal state or a proprioceptive prediction error always results in an immediate movement, this is not always the case. A simple demonstration is in delayed reaching tasks where motor intentions are already present in the preparatory phase but the onset of action execution depends on a sensory cue that is presented at a particular time. The precisions modulation can then also be used to separate the two phases of action preparation and execution (see Fig. \ref{fig_arm}), as done in \cite{ps22a}. In short, a delicate balance is in place between high- and low-level precisions: indeed, movement in Active Inference is possible through sensory attenuation, i.e., by reducing the precisions of sensory generative models, so that the belief can be free to change by its own dynamics, ultimately affecting what sensations will be sampled next \cite{bapef13}.

A more realistic implementation of reaching (and other advanced) movements is illustrated in \cite{Priorelli2023b}. This study introduces a block called \emph{IE model} that consists of intrinsic (e.g., joint angles) and extrinsic (e.g., Cartesian positions) beliefs, linked through a generative kinematic model. While the previous examples required inverse models (either in the dynamics function -- as a pseudoinverse or a Jacobian transpose -- or in the implicit backward pass of the VAE), in \cite{Priorelli2023b} the inversion arises naturally, through inference. The extrinsic goal is defined at a lower level compared to the intrinsic state, following the causal relations encoded into the generative process. Maintaining an extrinsic belief has critical benefits, since it permits to easily design complex movements (e.g., circular or linear trajectories), without worrying about intrinsic transformations. Furthermore, different blocks can be combined to encode intrinsic and extrinsic information for each DoF of the kinematic chain separately. This scheme affords more efficient control, permitting to simulate whole-body kinematics during (for example) sophisticated reaching and obstacle avoidance tasks. See Figures \ref{karate_kid} and \ref{matrix} for some examples.

Beyond reaching, Active Inference has been used also to simulate the control of eye movements \cite{paf14, pf19,pf18b}. For example, the study of \cite{Priorelli2023c} uses a hierarchical model that includes a belief over the target in absolute coordinates and a belief over the vergence-accommodation angles. The model consists of two parallel pathways -- one for each eye -- that perform a perspective projection of the target into the eye planes. This approach permits: (i) inferring the depth of the target by averaging the contributions from both eyes; (ii) fixating a target, by imposing an attractor in the projective space of the eyes; and (iii) performing concurrent depth estimation and target fixation -- or active vision -- through action-perception cycles -- which is particularly useful to counteract the nonuniform fovea resolution of the eyes (see Figure \ref{frames}).

Taken together, the above examples show that in Active Inference, it is possible to define goals for movement as priors over the internal representation of the system dynamics. These priors then lead to a goal-directed control through Free Energy minimization, rather than appealing to stimulus-response mappings and cost functions as in Optimal Control \cite{tj02, dsi10} and reinforcement learning \cite{sb98} (see \cite{f11} for a detailed discussion of the differences between the roles of cost functions in Optimal Control and Active Inference). As explained before, priors play a similar role as setpoints for movement control in cybernetics. The above examples also help illustrate the differences between the generative process encoding the real environmental dynamics, and the agent's generative model -- and the fact that they are reciprocally connected in an action-perception loop. Finally, they illustrate that there are various ways in which movement can be generated, e.g., it is possible to impose different kinds of priors that determine different behaviors; we will return to this point in the Discussion.


\subsection{Active sensing}

\begin{figure*}[!t]
    \centering
    \includegraphics[width=0.77\textwidth]{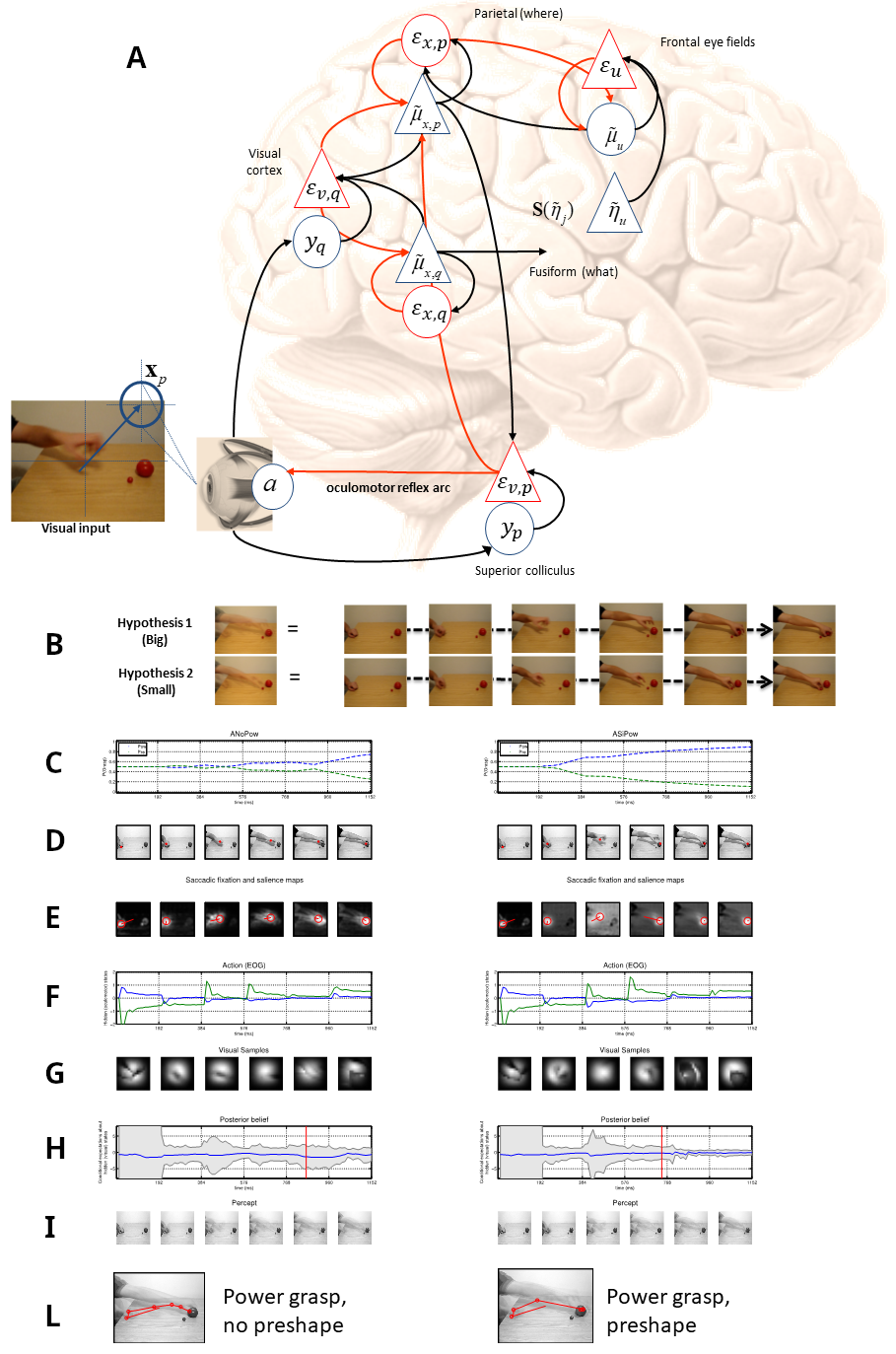}
    \caption{The active sensing model of \cite{dcafp17}. (A) The model describes which visual stimuli should be expected under two perceptual hypotheses (e.g., if the action target is the big/small object, when doing a saccade to the next hand position I should see a power/precision grasp) and generates saccades to check if the expectations are correct and to revise the probability of the two hypotheses. In the first hierarchical layer of the architecture, proprioceptive and visual signals $y_p$ and $y_q$ are generated, which are then used to compute the respective expectation $\tilde{\mu}_{x,p}$ and $\tilde{\mu}_{x,q}$ through message passing of prediction errors $\varepsilon_{x,p}$ and $\varepsilon_{x,q}$. (B) Schematic illustration of the two competing hypotheses, corresponding to sequences of images. (C-L) Simulation results of two representative trials, during which the agent observes an actor that grasps the small object without hand preshape (left) or with hand preshape (right); see the main text.}
    \label{fig5}
\end{figure*}

Active sensing refers to the ability of an agent to adapt its perception using self-generated energy to sample the environment \cite{aa16, mhks15, zh20}. Several Active Inference models implement active sensing routines to support visual processing \cite{ppf22, fapb12} and whisker movements \cite{mmbp21}, among other examples.

The model of \cite{dcafp17} combines motor prediction -- the reuse of the motor system to predict perceived movements -- and an active sensing (or hypothesis testing) strategy: the use of saccadic eye movements to disambiguate among alternative hypotheses. The architecture embeds a generative model of how (arm and hand) actions are performed to generate hypothesis-specific visual predictions, and directs saccades to the most informative (or diagnostic) visual locations to test them. The model follows the hierarchical form for generalized predictive coding, as shown in Fig. \ref{fig5}A. Internal states encode a representation of the center of oculomotor fixation and the probability that each hypothesis is the cause of the visual input. Hidden controls determine the location that attracts the gaze. The model is tested by evaluating the salience of sampling dynamic visual locations under two competing hypotheses (power grasp versus precision grip) in two conditions: with and without preshape. 

The model reproduces the differences observed empirically between the observation of goal-directed grasping actions, with or without informative cues (i.e., when the hand of the actor is ``preshaped" to grasp one of the two possible objects, big or small, versus when there is no preshape). The study of \cite{acs11} reported that during the observation of goal-directed grasping actions without informative cues (e.g., without preshape), visual saccades tend to follow the observed arm. Rather, when informative (preshape) cues are present, people make anticipatory saccades to the object to be grasped.  

The simulation results show significant differences between a reactive hand-following gaze strategy, which emerges in the no-preshape condition, and an anticipatory gaze strategy, which emerges in the preshape condition, shortly after the beginning of a trial - analogous to the empirical study of \cite{acs11}. The crucial model component that affords this active sensing strategy is a saliency map that assigns salience to the elements of the visual scene that afford information gain about the to-be-inferred grasping movements (hand shape) or their destination (objects) - in such a way that the objects only become salient when the uncertainty about the grasping movement has been resolved. See \cite{dcafp17} for details.

The authors assume that a hierarchically organized “action observation” brain network computes both the expected hand position (at lower hierarchical levels) and the probability of the two competing hypotheses (at higher hierarchical levels). Fig. \ref{fig5}B shows the two competing hypotheses considered, which are not only about final states (hand on big versus small object), but encompass the whole action unfolding in time. In practice, they correspond to sequences of (superimposed) images of hand trajectories (here, 6 time frames). As evident in the figure, the hypothesis that the actor is reaching a small (or big) object entails that the hand will be configured in a precision grip (or power grasp) during action execution -- and it is this sort of hypothesis that the agent tests by performing saccades to the most informative locations. Fig. \ref{fig5}C-L shows the results of two example trials during which the agent observes an actor performing a precision grip toward the small object, without (left) or with preshape (right). In particular, the panels show the expected probability of the two competing hypotheses during an example trial (C); the location of the saccades in the video frame at six time frames (D); the corresponding saliency maps, with white locations corresponding to the locations to which the model assigns greater salience, hence the best candidates for the next saccade (E); the hidden (oculomotor) states computed by the model (F); the content of what is sampled by a saccade in the (filtered) map (G); the posterior beliefs about the “true” hypothesis, where expectations (expected log probabilities) are plotted in blue and the associated uncertainty (90\% confidence interval) in gray (H); the observations of the model i.e., the mixture of the viable hypotheses weighted by the posterior expectation, represented as a weighted superposition of all the frames during the simulation steps (I); and the sequence of saccades that the model performs during the experiment (L). 

Without preshape (left panels), the gaze follows a reactive, hand-following strategy and the action is disambiguated fairly late in the trial. Rather, with preshape information (right panels), the clues present in the hand movement afford a faster disambiguation of the correct hypothesis and anticipatory saccades to the inferred objects: the eyes land on the small object before the hand arrives. This example illustrates that the same Active Inference mechanism that we discussed above in the context of goal-directed actions (e.g., reaching an external target) can also model active and higher-level oculomotor control, where the objective is to sample the sensorium for hypothesis testing. Both forms of behavior stem from Free Energy minimization under different generative models; while we illustrate them separately, they could also emerge simultaneously in the same model \cite{ppf22}.


\subsection{Unintentional actions driven by multisensory conflict}

So far, we discussed Active Inference models of goal-directed movement (e.g., reaching or following a target) and active sensing. However, movement can also arise unintentionally, with little or no awareness. While the study of unintentional motor behavior found so far little space in the motor control literature, recent evidence for the systematic induction of unintentional actions comes from embodiment studies, in which subjects undergo an illusory experience in which fake bodies (e.g., virtual avatars) or body parts (e.g., rubber hands) are perceived as being (part of) their own body \cite{bc98, esp04, ms13, pe08}. During these body ownership illusions, subjects process the seen body (parts) as the same causal entity generating somatosensory sensations \cite{cem22, kmks15, scs15}, and to some extent it is possible to introduce multisensory conflicts about the body configuration without breaking the illusion \cite{ll07, ma16}. For example, in the Rubber Hand Illusion (RHI), when the rubber hand is placed next to the real (occluded) hand, a visuo-proprioceptive conflict about the hand location is in place. This conflict has been associated with a proprioceptive recalibration of the perceived hand location since the very first report of the RHI \cite{bc98}, and consists of a shift of perceived hand location in the direction of the visual hand. Interestingly, later works have associated an active component to the illusion; namely, the tendency to unconsciously exert a force in the direction of the visual hand and to move along with it if no restrictions are in place, in some cases even when subjects are explicitly instructed to stay still \cite{lfmf21, a15, gcbm20}. This behavior has been associated with an active strategy for suppressing prediction errors associated with the perceived location of the hand. Additionally, movements may arise because the subject tries to minimize model uncertainty, e.g., to understand if the perceived and the real hand positions match. The latter process has been successfully reproduced in Active Inference implementations of the action-perception loops during body ownership illusions \cite{mlp22,lfmf21,gcbm20}.

\begin{figure*}[!t]
    \centering
    \includegraphics[width=0.75\textwidth]{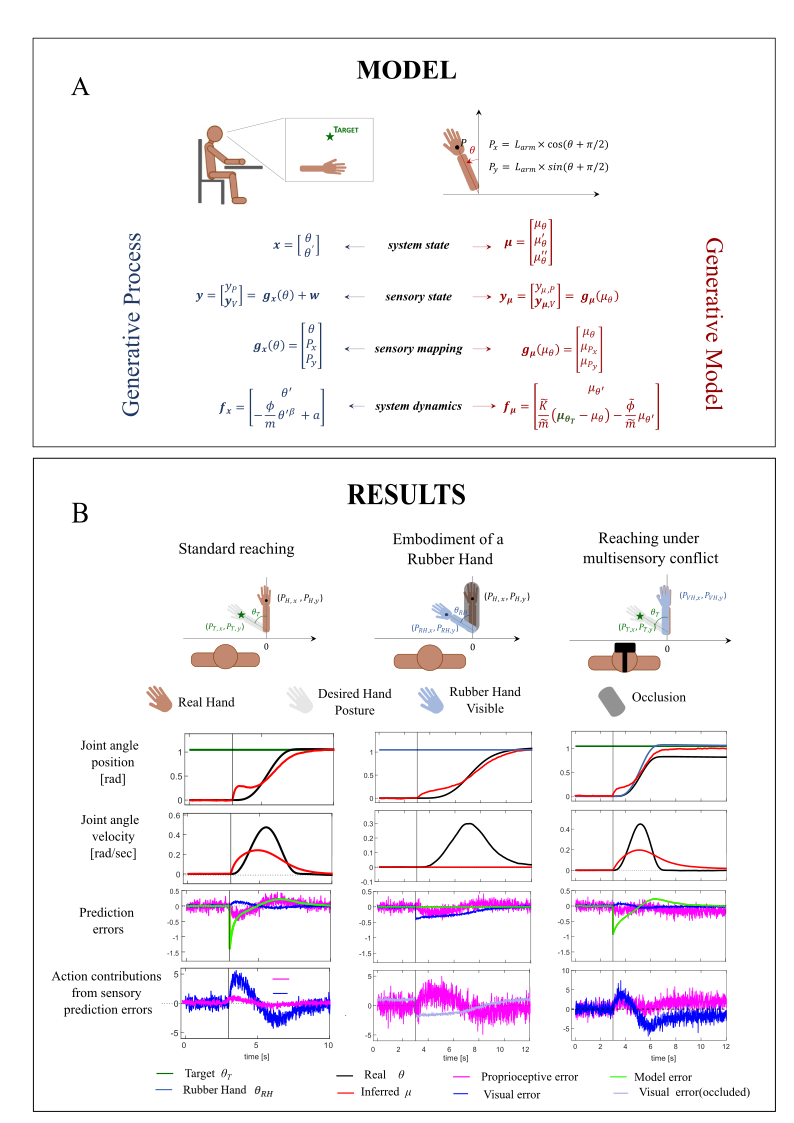}
    \caption{Schematic implementation and results from the Active Inference model that simultaneously accounts for intentional (goal-directed) and unintentional movements, described in \cite{mlp22}. (A) The model implements a 1-DoF agent, whose configuration is uniquely described by its elbow’s joint angle and angular velocity -- $\boldsymbol{x} =[\theta, \theta^\prime]$ -- and who receives information about its own configuration and the environment through proprioception and vision -- $\boldsymbol{y} =[y_P, y_V]$. The dynamics of the real arm -- $\dot{\boldsymbol{x}} = f_x(\boldsymbol{x},\boldsymbol{a})$ -- represents a damped system that may be subject to internal forces generated through actions (here, formalized as joint angular accelerations), while the internal model of the arm dynamics $f_\mu(\mu_\theta, \mu_{\theta_T})$ by a damped oscillator where the attractor is either set to the arm configuration $\mu_{\theta_T}$ in which the hand is on target (for reaching actions), or to the current state when the agent has no intention to move. (B) Results from three simulations: standard reaching (left column), classic rubber hand illusion (central column), and reaching under visuo-proprioceptive conflict (right column). In each column, panels from top to bottom show the temporal evolution of the real and inferred joint angles and joint angular velocity, of the prediction errors, and of the contributions to action arising from the minimization of proprioceptive and visual prediction errors.}
    \label{fig6}
\end{figure*}

The study of \cite{lfmf21} introduced an Active Inference model of the active strategies that emerge during ownership illusions to suppress multisensory (self-perception) conflicts. The model is tailored to the classic RHI in its virtual version (using a virtual hand, not a rubber hand), in which subjects are not allowed to move their hand. In line with this, the model computes the actions but these do not enter the computation of the hand dynamics. Indeed, actions are here computed as a proxy for the force that participants exerted while experiencing the illusions with their arms restrained. In this respect, the model has the intrinsic limitation of being uncoupled with the arm dynamics. This limitation was addressed in another study \cite{mlp22}, which proposed a unified model that can account for both goal-directed motor behavior (i.e., reaching actions) and unintentional motor adjustments arising from multisensory conflicts associated with self-perception, and for their interaction. 

The model of \cite{mlp22} implements an agent that continuously infers its own bodily configuration and could be set to have a goal of reaching a given target. If no goal to reach is instantiated, which corresponds to no intention to move, the agent is set to fulfill the requirement to keep its current configuration. A schematic summary of the model is given in Fig. \ref{fig6}A. An important novelty with respect to previous implementations of arm control is the possibility to simulate the case in which the agent has no goal (i.e., no intention to move), enabled by keeping the internal representation of system dynamics used for reaching tasks (i.e., a damped oscillator) and setting the attractor (i.e., the desired state) to the current arm configuration. Despite its simplicity, this extension is key, because it allows inspecting subtle aspects of movement such as how unintentional motor adjustments arise as a byproduct of self-perception, and how these adjustments interfere with goal-directed behavior. This is exemplified by some of the results of the model that we discuss in the following. A second difference from most previous implementations is that action is computed by concurrently minimizing prediction errors in the proprioceptive and the exteroceptive (visual) domains. This is essential to correctly model visually-guided action operated under multisensory conflict about the body state.

Fig. \ref{fig6}B shows results comparing standard reaching, unintentional alignment of the physical hand to its visual counterpart during an ownership illusion, and reaching under bodily multisensory conflict (columns from left to right respectively). The three simulations use the same agent exposed to different combinations of tasks and sensory inputs. In the first, the agent is assigned a standard reaching task; in the other two, the agent undergoes an ownership illusion over a rubber/virtual hand, and sensory input about limb state is streamed by the fake the real hand in the visual and the proprioceptive domain, respectively. In one case, the static rubber hand is displaced with respect to the real hand and the agent does not have a task assigned besides observing (and inferring) its own state. In the other case, the agent has to reach a target but the velocity of the virtual hand is set to 1.3 times the one of the real hand, so that during the task execution the two hands get progressively displaced from one another. The model assumes that the ownership illusion is in place by treating the visual input from the fake hand as if generated by the inferred arm configuration, which is encoded by the agent in the proprioceptive domains ($y_\theta$). Thus sensory predictions takes the form:  $\boldsymbol{y}_\mu = [ y_{\mu_P},  \boldsymbol{y}_{\mu_V}]$, with $y_{\mu_P} = \mu_\theta$ and $\boldsymbol{y}_{\mu_V} = [\mu_{P_{RH, x}},  \mu_{P_{RH, y}}]$ (see Fig. \ref{fig6}A for more details). In the simulation of the static RHI,  the internal model has been adapted by tuning two of its parameters: (i) the gain of the action component driven by vision, set to zero to account for the fact that the fake hand is static and not under the control of the agent, and (ii) the internal estimate of the sensory noise in the visual domain, increased to roughly account for the fact that (as it happens in the real world) while undergoing an ownership illusion the agent is still aware that the seen hand is fake and therefore “less reliable” as a source of information about its own bodily state; see \cite{mlp22} for more details.

The results from the three simulations shown in Fig. \ref{fig6}B demonstrated that the proposed implementation could account for intentional motor behavior, as in the exemplification of standard reaching, and for unintentional motor adjustments driven by multisensory conflict in self-body processing, as observed in experimental settings of ownership illusions. In addition, the model successfully reproduces the motor behavior observed for visually-guided actions in presence of multisensory conflicts, like in the case of reaching under visuomotor rotations or with aberrant velocity mappings. Importantly, this case would indicate that the action is driven by both visual and proprioceptive prediction errors; in fact, keeping action exclusively associated with the minimization of proprioceptive prediction errors would lead to the implausible result of the agent overshooting the visual target (see \cite{mlp22} for details). Interestingly, a previous Active Inference model of a similar visuomotor rotation task has shown that modulating the relative weighting of the internal estimates of the (visual and proprioceptive) sensory noises can mimic attentional effects akin to those observed in the laboratory \cite{lim20}.

Together, these results lead to two main insights. First, in the case of intentional reaches, prediction errors are initially dominated by model errors driven by the internal dynamics under the effect of the target attractors; then, sensory prediction errors arise as a consequence of the model errors on perceptual inference. In the case of the RHI, model errors are absent (as no motor intention is instantiated) and the sensory prediction errors -- thus the actions -- arise as a byproduct of the visuo-proprioceptive conflicts associated with self-body perception. An important consequence of the lack of model errors is that the agent does not update the internal estimate of the joint angular velocity, which stays null as if no movement has occurred (providing that the agent has no direct access to joint velocity through proprioceptive receptors). This has been suggested as a possible explanation for the lack of motor awareness that typically characterizes subtle unintentional motor adjustments.  

Second, when simulating reaching under visuo-proprioceptive conflicts, a better fit with experimental data can be obtained by also allowing exteroceptive (not only proprioceptive) prediction errors to drive the action. Given the spatial misalignment that emerges from the aberrant velocity mapping between real and fake hands, the inferred posture ($\mu_\theta$) is biased toward the visual hand. If action were only driven by proprioceptive prediction errors, both the virtual and the real hand would overshoot the target, which runs against the empirical observation that the task is accomplished once the visual hand correctly reaches the target.


\subsection{Mixed models for sensorimotor decisions}

Despite the traditional literature views decision-making and action control systems as separated cognitive processes, a more recent tendency considers them as two interacting levels of the same integrated system \cite{gcwf18, lp15, pc16}. Accomplishing various skilled tasks engages a series of decisions to establish the sequence of movements to make, and to guide the sensorimotor behavior. At the same time, actions can determine changes in the real world that force us to modify the goal of the executed task and to revise the plan previously imagined.

From a modeling perspective, the variables employed in this sensorimotor decision-making system are of different natures: decision-making typically involves discrete variables that select the sequence of actions composing a motor behavior, while the execution of motor behavior induces a dynamic variation of some continuous variables (e.g., contracting muscles or decreasing the body temperature). To integrate both discrete and continuous time variables within the same Active Inference model, it is possible to adopt so-called “mixed” models \cite{fpv17}. Mixed models inherit the architecture of hierarchical generative models, where the prediction of a level acts as a prior for the level below, which in turn computes a prediction error that is then used as a likelihood signal for the higher level. In a typical mixed model having two layers, the higher layer consists of a discrete Partially Observable Markov Decision Process (POMDP); in this article, we did not focus on this discrete-time Active Inference, but a full treatment can be found in \cite{ppf22}. Rather, the lower layer of the typical mixed model implements exactly the Active Inference in continuous time that has been the focus of this article. The higher layer generates sequences of discrete outcomes, which constitute the priors – or fixed-point attractors – on the hidden causes, guiding the sensorimotor process controlled by the lower layer. Since the two layers consist of variables of different natures, they are connected by a particular interface described in \cite{fpv17} that propagates the beliefs from one layer to another, via descending and ascending messages.
 
In mixed models, inference progresses by determining probabilities $\boldsymbol{\pi}$ about sequences of control states (i.e., policies) at the higher layer. Each policy $\boldsymbol{\pi}_\pi$ generates a transition between hidden discrete states $\boldsymbol{s}_{\pi,\tau}$, which corresponds to a sequence of predicted outcomes $\boldsymbol{o}_{\pi,\tau}$. By performing a Bayesian model average of the outcomes of all policies, a posterior predictive distribution $\boldsymbol{o}_\tau = \sum_\pi \boldsymbol{\pi}_\pi \cdot \boldsymbol{o}_{\pi,\tau}$ is obtained and sent as a descending message to the lower continuous layer. Each component $o_{\tau,m}$ can be understood as a particular model, steering the lower-level continuous dynamics in a specific direction. The mapping of an outcome model into the continuous space is denoted as $\boldsymbol{\eta}_m$, encoding a fixed empirical prior; a second Bayesian model average defines the actual prior $\boldsymbol{\eta}$ over the hidden causes $\boldsymbol{\nu}$, i.e., $\boldsymbol{\eta}= \sum_m \boldsymbol{\eta}_m \cdot o_{\tau,m}$. Having sampled continuous observations, the lower layer returns an ascending posterior estimation for the belief of each discrete outcome through a model evidence accumulated by the dynamical system over some time $T$:
\begin{equation}
    \mathbf{E}(t)_m = - \ln o_{\tau,m} - \int_{0}^{T} \mathbf{L}(t)_m dt    
\end{equation}
where $\mathbf{L}(t)_m = \ln P(\tilde{\boldsymbol{y}}(t)|\boldsymbol{\eta}_m)-\ln P(\tilde{\boldsymbol{y}}(t)|\boldsymbol{\eta})$, with $\ln P(\tilde{\boldsymbol{y}}(t)|\boldsymbol{\eta}_m)$ and $\ln P(\tilde{\boldsymbol{y}}(t)|\boldsymbol{\eta})$ denoting the log evidences about continuous observations  $\tilde{\boldsymbol{y}}(t)=[y(t),y'(t),y''(t),\dots]$ regarding a single model $m$ and the full set of models, respectively. In other words, $\mathbf{L}(t)_m$ is a post-hoc Bayesian comparison between two (Gaussian) probability densities used to sample the outcomes, under the empirical reduced ($\boldsymbol{\eta}_m$) and the full ($\boldsymbol{\eta}$) priors \cite{fp11}. It can be shown that if $\boldsymbol{\eta}_m =\boldsymbol{\eta}$, then $\mathbf{L}(t)_m=0$; see \cite{fpv17} for a demonstration.
 
On the other hand, $\mathbf{E}(t)_m$ expresses the Free Energy of competing outcome models defined as the sum between their descending prior surprise $-\ln o_{\tau,m}$ and the log evidence $\mathbf{L}(t)_m$ of their ascending posterior integrated over time. Note that when $T=0$, the ascending posterior reduces to the descending prior. Thus, $\mathbf{E}(t)_m$ assigns a score to the sampled continuous outcomes in relation to the predicted discrete models. To convert this score for each model back to the discrete layer, $\mathbf{E}(t)$ is passed through a softmax function to give a posterior over each outcome model, so that it can be used as discrete observation into the POMDP inference process.
 
In the last few years, various studies have used mixed models to target scenarios requiring both discrete and continuous variables. For example, \cite{pf18} proposed a mixed generative model to sample visual information from the environment, which shares some resemblances with the active sensing model of \cite{dcafp17} introduced above, but integrates both discrete and continuous variables. The discrete layer of the model implements a POMDP to build a sequence of saccade targets and to decide where to look. Such decisions are translated into movements of the oculomotor system by the continuous layer that implements how to look, i.e., the realization of the saccades by controlling the anatomical effectors. Successively, the sampled observations are fed back to the discrete layer to evaluate the goodness of the saccade sequence.
 
A similar mixed generative model was used to study the interaction between pharmaceuticals and oculomotor behaviors, by focusing on the influence of cholinergic and GABAergic agents upon the choice of the target to fixate and the speed of saccades \cite{pf19}. The authors simulated an oculomotor task introduced in \cite{fbg89}, in which a cue for a given saccade location is presented and after its disappearance, a saccade to the target is executed. In the mixed model used to simulate the oculomotor task, the discrete layer generates predictions about the fixation locations, which constitute the attractor points for the oculomotor system dynamics in the continuous state-space. The effects of neuromodulators are simulated by changing various parameters of the mixed model; namely, the precision of the hidden states transitions (noradrenaline), the mapping between hidden states and sensory data (acetylcholine), the belief about the best saccade to select (dopamine), or the empirical priors that control the saccade peak velocity (GABA).
 
Another application of mixed models is the “active listening” model \cite{fsqpph21}, which simulates the parsing of meaningful words from auditory perception. Following some insights borrowed from active vision, the generative model segments the continuous stream of acoustic signals by placing word boundaries in accordance with some prior constraints. For example, the offset of one word should precede, in some plausible time range, the onset of the subsequent word; in a speech, signal segmentations are more likely to contain words than non-words; chosen a specific language, there exists a prior knowledge on the possible produced words; etc. The active listening model then proceeds by identifying several plausible boundary intervals, which provide the greatest evidence for the prior beliefs about the words.  

Another interesting application is illustrated in \cite{Parr2021}. This study shows that Active Inference can simulate several neurological conditions and that some reflexes might naturally emerge under the appropriate generative model. Furthermore, this implementation shows that linking a continuous model of the arm dynamics and a discrete decision model permits performing multi-step reaching movements, by planning from a high-level goal. A similar model was used to solve a dynamic pick-and-place operation \cite{Priorelli2023d}. In this case, two novelties are introduced. First, the discrete model generates and integrates predictions simultaneously from the intrinsic and extrinsic modalities also used in \cite{Priorelli2023b}. Second, the reduced priors of the agent are updated at each discrete step, allowing it to grasp moving objects. A similar grasping task was simulated in \cite{Priorelli2023e}, but using an unconventional approach. In this case, the hidden causes were sampled by a categorical distribution, while the reduced priors were generated by independent dynamics functions of the hidden states. This allowed to impose and infer static and discrete intentions corresponding to dynamic trajectories in the continuous environment.
  
Yet another example of mixed models - this time, in the domain of interoceptive processing and autonomic (not action) control - is provided by \cite{tbmbsp21}. The authors describe the mechanisms of adaptive physiological regulation using three generative models of increasing complexity, which are able to simulate homeostatic, allostatic, and goal-directed regulation of bodily and interoceptive parameters, such as temperature, thirst, and hunger. While the first two generative models use only continuous variables, the latter generative model (for goal-directed control) is a mixed model, in which the higher layer implements a POMDP process to select among discrete policies (e.g., to run with or without a bottle of water), while the lower layer is a continuous time system that regulates interoceptive data (e.g., body temperature) via autonomic reflexes (which might be considered largely analogous to motor reflexes, but operate on interoceptive streams \cite{sf16, bs15}). In order to estimate the long-term consequences of a certain policy, the model maps discrete outcomes at the higher layer into prior beliefs for specific interoceptive observations at the lower layer. Conversely, the lower layer model provides evidence about the discrete outcomes used as hypotheses about the expected prediction errors, so that it contributes evaluating the policies at the higher layer. Models of this kind can be used to support computationally guided investigations of interoceptive processing and its dysfunctions that are possibly associated with psychopathological conditions \cite{bp18, fsmd14, a18, mbbp21, pfk19, pmbb19, bwpb17}.


\section{Discussion}

How does the brain control movements toward goals? There is a view of motor control -- pioneered by ideomotor theory and cybernetics -- according to which actions are inextricably linked to their effects, rather than stemming as responses to stimuli. In these theories, action starts with some internal image of an intended effect -- sometimes called a preference, a goal, or a setpoint -- and the movement is the consequence of filling the gap between the intended effect and the sensed environmental condition. In other words, these theories assign a role to action effects, or the discrepancy between action effects and sensory events, in the selection and the control of movements. Active Inference formalizes key intuitions of these theories, in terms of priors, predictions, and prediction errors, therefore linking to a large body of studies about predictive processing and Bayesian inference in biological organisms \cite{pc16, h13, ppf22, p17, wmco20} and in robotics \cite{a23, cvvds21, csphl21, a21, mt20, mv22}.

Here, we provided a brief illustration of Active Inference in continuous time and discussed specific models that targeted various aspects of motor control; namely, the execution of goal-directed reaching actions, active sensing, the resolution of multisensory conflicts, and the integration of discrete (decision-related) and continuous (perception- and action-related) processes. Each of the example models that we have briefly reviewed can be evaluated by its own merits, such as by its capability to accurately account for empirical data. However, taken together, these models (and others) show that Active Inference can address a large variety of motor control processes. Importantly, all the motor control phenomena illustrated by our examples stem from the same process of Free Energy minimization, rather than requiring separate objective functions. This feature makes Active Inference appealing both as a general theory of biological systems and as a technical framework to advance AI and robotics research. 

Despite the appeal of the models that we have reviewed, Active Inference accounts of motor control are still relatively young compared to other frameworks, such as Optimal Control \cite{tj02, dsi10}. Several open issues need to be clarified to develop Active Inference accounts of motor control that are more mature from both biological and robotic perspectives. Below we briefly discuss some of the most important open issues that need to be addressed in future research.

One open issue concerns the sensory modalities involved in motor control. As discussed in Section 2.3, some biological considerations suggest that motor control could be realized by minimizing proprioceptive, not exteroceptive prediction errors \cite{asf12}. However, as highlighted in \cite{mlp22}, driving action with exteroceptive errors would seem necessary to correctly reproduce visually-guided reaching behavior in the presence of visuomotor conflict. From a practical perspective, including exteroceptive modalities in the Free Energy minimization through action offers various advantages \cite{olc22, sgl20}. For example, the reaching model of \cite{fdkk10} uses proprioceptive and visual sensory modalities for perception and action. An advantage of this approach is that the agent can perform smooth and accurate movements even in the presence of high proprioceptive noise, given that the visual input is more stable. As noted in \cite{ps22a}, the increased stability derives not only from the fact that there is less noise in the action update, but also because both the action and the high-level belief are updated with the same information, and the effect is more prominent as the visual precision increases. Further studies are thus needed to understand the actual role of visual predictions in action execution or, more clearly, how visually-guided movements can be correctly realized, from a biologically plausible perspective, in presence of noise or conflicts between different sensory modalities.

This open issue implies that there might be a tension between standard formulations of Active Inference that focus on biological aspects, and studies that realize efficient robotic implementations.
Besides multisensory integration, an Active Inference agent may also act by minimizing increasing temporal orders of the prediction error, as in \cite{bdlh21}, where an agent is controlled by both position and velocity, resulting in increased stability and additional control over the environment, if an appropriate attractor is embedded at high orders of the belief dynamics. 
Design differences exist also about the temporal order primarily affected by the attractor: while this is usually embedded into the 1st-order dynamics function, some models encode it in the 2nd-order to achieve a more stable control, especially when the robot is force-controlled \cite{pnfp16, fdk09}. Finally, different models use different kinds of errors as the attractive force for motor control. As discussed in Section 2, the belief update depends on three components: a likelihood error from lower hierarchical levels, a backward error coming from the next temporal order, and a forward error from the previous order. Generally, the attractive role is fulfilled by the backward error, which however requires the computation of the gradient of the dynamics function \cite{sgl20}. Other studies use instead the forward error (which is simpler to compute) as the main attractive force \cite{ps22a}. Finally, an alternative strategy consists of including control costs in the Free Energy expression, to remove estimation biases and afford optimal action \cite{baioumy2022unbiased}.
The pros and cons of the different approaches and their biological plausibility remain to be systematically investigated.

Another dimension that is important to consider in Active Inference studies is the way the generative model is designed or learned - since the generative model implicitly defines the agent's behavior. One crucial design choice regards the extent to which the generative model and the generative process are similar or dissimilar. For the sake of simplicity, many Active Inference studies use generative models that are almost identical to the respective generative processes, with few quantitative differences. In these studies, the generative model is usually aligned to the generative process, in three ways. First, the internal state variables are modeled as explicit representations of features of the physical environment or the body, so that the generative model already incorporates explicit task-related variables such as speed, pressure, position in the allocentric space, etc. Second, the internal prior dynamics are designed as a copy of the simulated world dynamics, in the sense that the sets of differential equations implementing the changes of the state variables are the same. Third, motor commands are built as inverse models of physical world/body features, so that actions are direct changes in speed, pressure, positions in the allocentric space or other physical entities. For example, \cite{bapef13} shows an Active Inference model of the behavior in a force-matching task, where subjects have to match a reference force by pressing directly on themselves. In this case, physics is simulated with two coupled differential equations defining the dynamics of the self-generated and the external force. Sensory (proprioceptive and somatosensory) observations are simple linear mappings of these hidden variables with the only exception that while proprioception is a linear mapping of self-generated forces alone, touch is a raw sum of self-generated and external force. The prior dynamics of the generative model are then built in strict relation to the generative process, as a set of differential equations which is quite similar to the one described above, with the only exception that a causal variable takes the place of the action. Accordingly, all mappings to the sensory predictions also have the same features as the ones generating the described observations from the simulated physics. A similar example of the similarity between generative model and generative process is offered by a model of the accommodation of delays in oculomotor control \cite{paf14}. In the model, the generative process consists of a set of ordinary differential equations, which describes the dynamics of the current oculomotor displacement and the target location; and the mappings generating sensory observations (displacement and target position) are linear combinations of the hidden variables. As in the previous example, the generative model is simply a copy of the generative process, except for the fact that the latter includes a contribution from the action.

Nevertheless, Active Inference does not necessarily require that the two systems are the same. What is important is that the generative model affords adaptive motor control, by translating the internal dynamics into commands to the motor actuators, to change the environment in a predictable way. This is in keeping with the ``good regulator theorem", which states that a good controller needs to either include or be (embody) a model of a system \cite{ca70, pdims17, s14}. One possibility is using generative models that only generate predictions at the level of the proximal (e.g., proprioceptive) features, which are the closest consequences of motor commands, rather than distal features. An example is a model of the active control of whisking behavior, in which the generative model only predicts the (somatosensory and proprioceptive) consequences of whisker movements. The model does not include any internal variable that directly represents the distance from external objects or their identity, yet it is able to estimate them implicitly \cite{mmbp21}. The estimation strategy is based on the active control of whiskers. Namely, whisker amplitude is continuously adapted to fit the (expected) distance to objects and at convergence, it could be used as an implicit inference of animal-object distance, as shown empirically \cite{aa16, vhc15}. Designing or learning appropriate generative models is a key prerequisite to accurately model motor control (or other) tasks. Current advances in machine learning permit inferring models from data, but it remains to be investigated how to better incorporate them into Active Inference models \cite{a23}.

Finally, another key issue that deserves further investigation is the link between the biologically motivated aspects of the theory and the computational models used in practical implementations e.g., deep neural networks. From a biological viewpoint, hierarchical Active Inference assumes a temporally deep model based on predictive coding, which uses local message passing of predictions and prediction errors across brain areas. In principle, an architecture of this kind would allow the formation of effective and increasingly more invariant representations of the sensory input at higher levels of the cortical hierarchy, via a biologically motivated scheme \cite{ok22}. However, in practical implementations, it is common to use (deep) neural networks as generative models \cite{sgl20} rather than hierarchical predictive coding. While using deep networks is effective, it does not take advantage of the local message passing of predictive coding such as in the hierarchical kinematic model of \cite{Priorelli2023b}. Furthermore, rather than assuming prediction error minimization at every level of the hierarchy, deep networks often only pass their final gradient to beliefs encoded at high hierarchical layers. A similar argument could be made for \emph{precision control}, which links to learning and attention in Active Inference. While the precision of signals at every level should be inferred by minimizing Free Energy, this is rarely done in practice. For example, in the studies illustrated in Chapter 3, the precision matrices of latent states were fixed. In principle, allowing Active Inference models to change the precision of signals at every hierarchical level should make them more adaptive and effective, but this possibility remains to be fully investigated in future studies.

\bibliographystyle{IEEEtran}
\bibliography{references}


\section*{Acknoledgments}
This research received funding from the European Research Council under the Grant Agreement No. 820213 to GP, the European Union’s Horizon 2020 Framework Programme for Research and Innovation under the Specific Grant Agreements 945539 to GP and 951910 to I.S., and the Italian Ministry for Research under Grant Agreements 2017KZNZLN to IS, 2020529PCP to FD and PE0000013-FAIR and IR0000011–EBRAINS-Italy to GP.

\newpage
\section*{Supplementary}

\setcounter{figure}{0}
\renewcommand{\thefigure}{S\arabic{figure}}
        
\begin{figure*}[!h]
    \centering
        \centering
        \includegraphics[width=\textwidth]{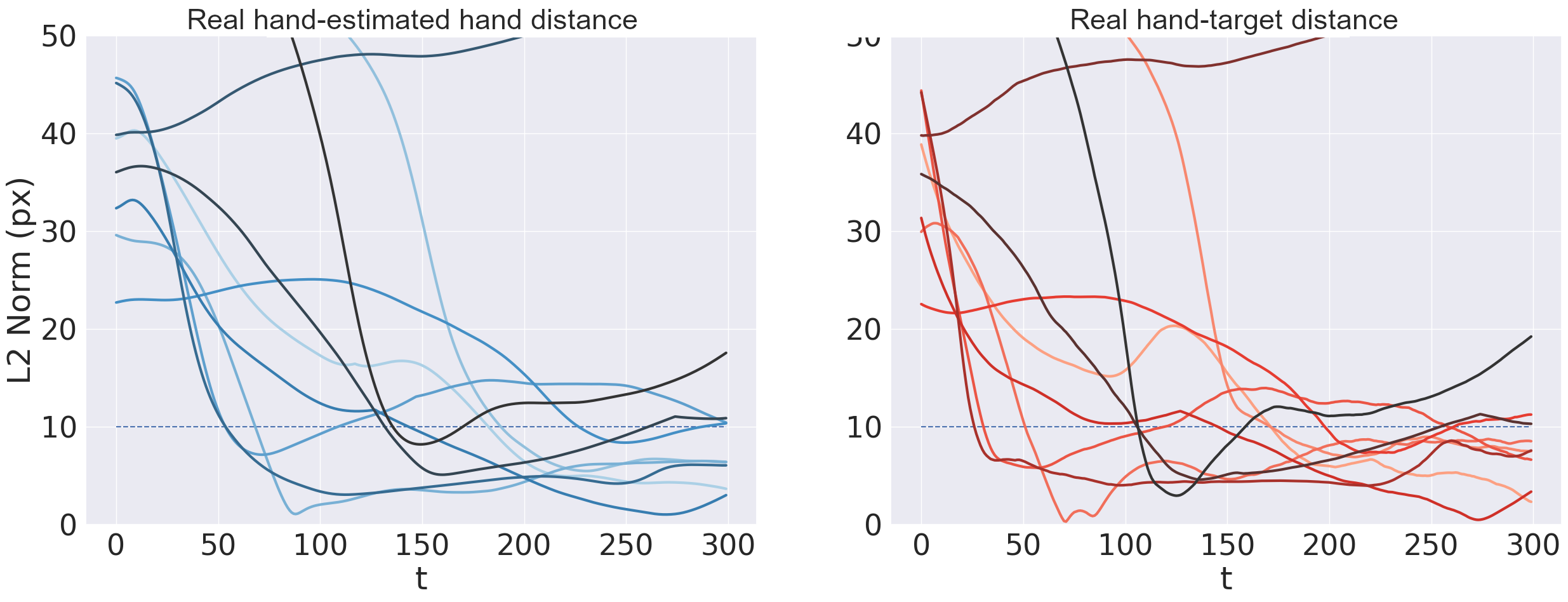}
        \caption{Illustration of the performance of the Active Inference model of \cite{ps22a} during a target tracking task. The left and right panels show the performance of the model during the reaching movement and target estimation. Each line corresponds to a trial. L2 distance between the real hand and target over time (left), and error between the real and estimated target positions over time (right). Both decrease in most trials, as the arm successfully reaches the target (the dotted line indicates the minimum distance from the target to consider a trial successful).}
        \label{fig4} 
\end{figure*}

\begin{figure*}[!h]
    \centering
        \centering
        \includegraphics[width=\textwidth]{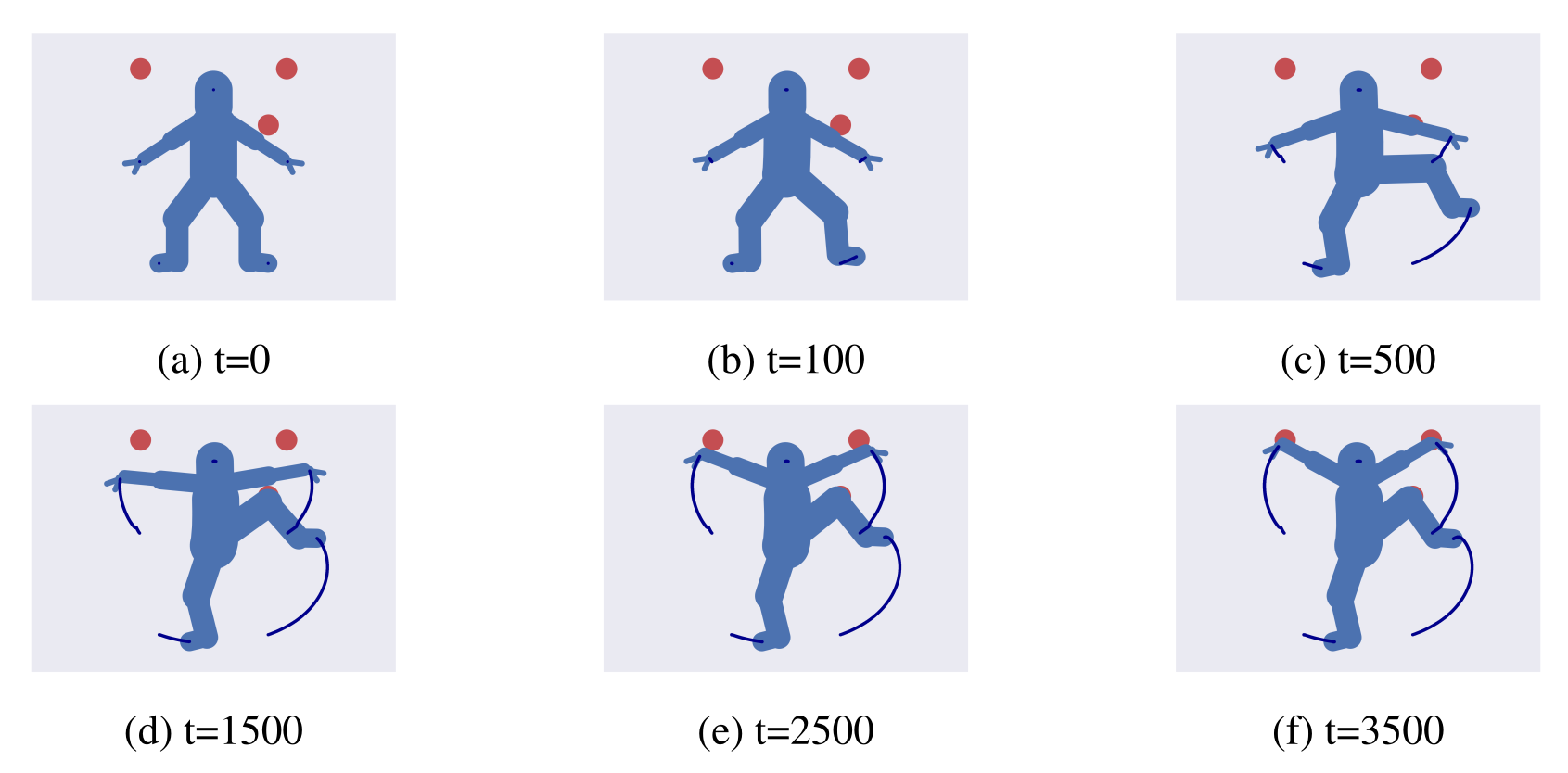}
        \caption{Controlling a simplified humanoid body composed of 23-DoF. The goal is reaching 3 different target locations, with the left knee and the two arms.}
        \label{karate_kid} 
\end{figure*}

\begin{figure*}[!h]
    \centering
        \centering
        \includegraphics[width=\textwidth]{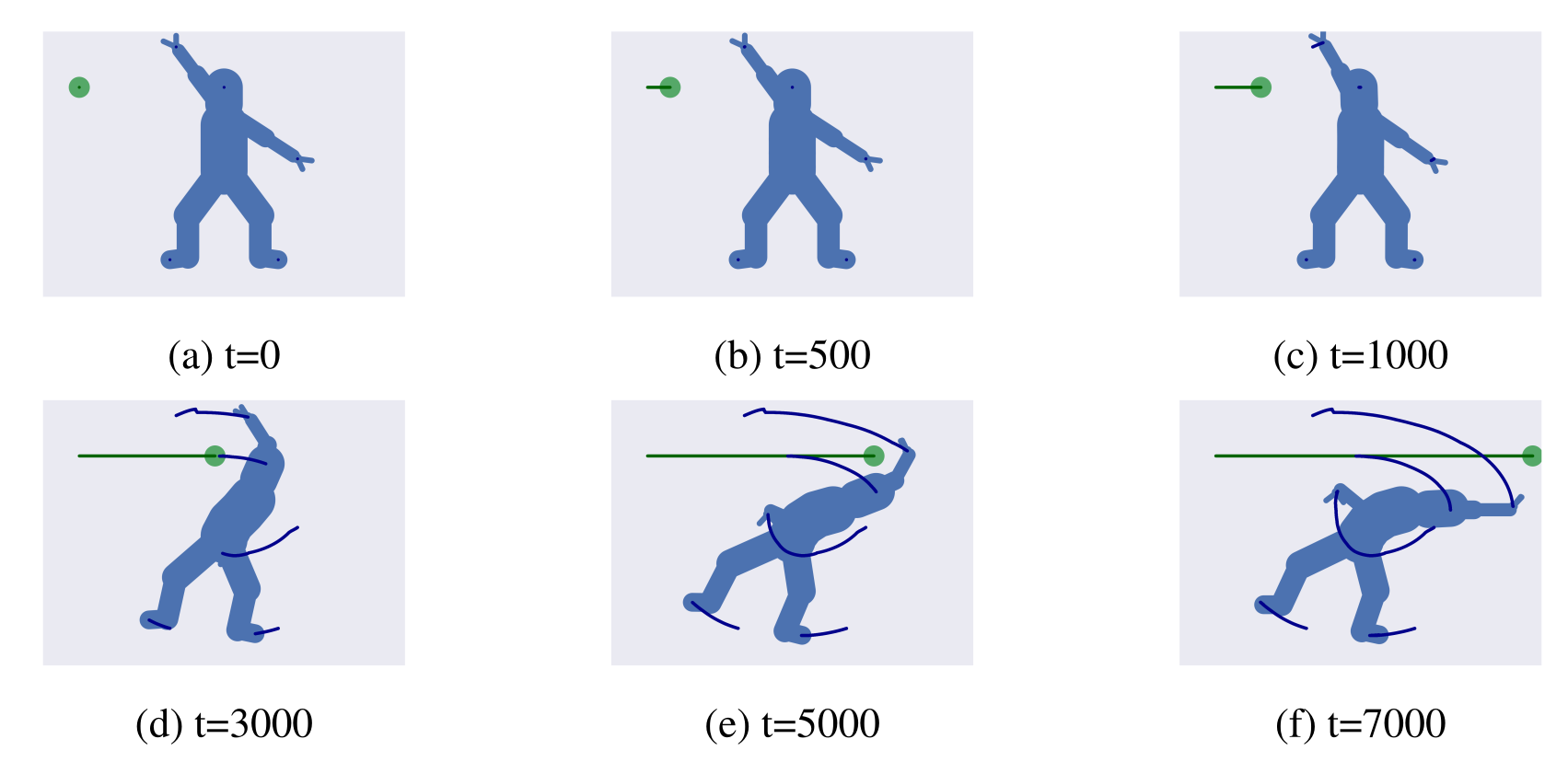}
        \caption{Controlling a simplified humanoid body composed of 23-DoF. The task consists of avoiding a dynamic obstacle with the whole body.}
        \label{matrix} 
\end{figure*}

\begin{figure*}[!h]
    \centering
        \centering
        \includegraphics[width=\textwidth]{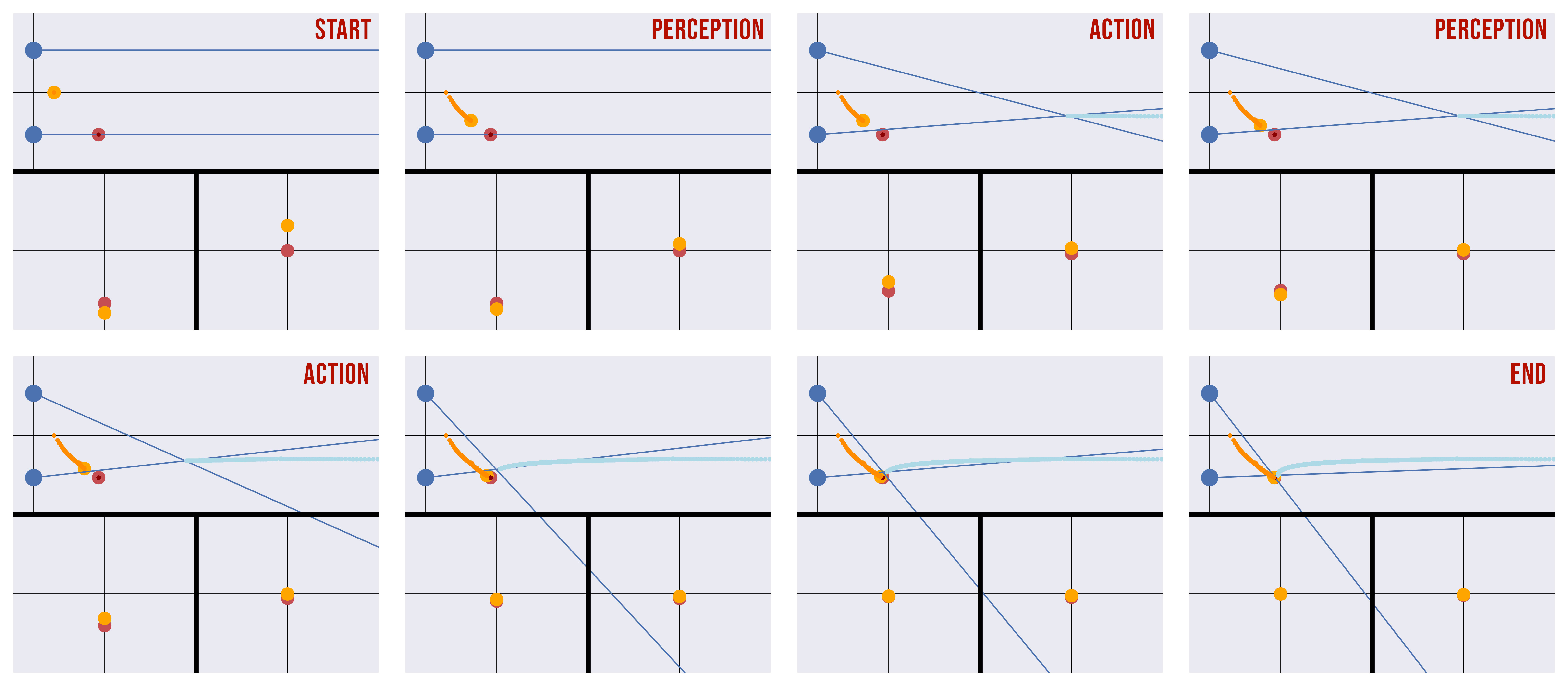}
        \caption{Sequence of time frames of a depth estimation task with simultaneous target fixation. The agent uses alternating action-perception phases to avoid being stuck during the minimization process. The eyes are represented in blue, and the real and estimated target position in red and orange. The fixation trajectory (when vergence occurs) is represented in cyan. Each frame is composed of three images: a view of the overall task (top), and the projection of the target to the two camera planes of the eyes (bottom left and bottom right).}
        \label{frames} 
\end{figure*}

\end{document}